\documentclass[twocolumn,tighten]{aastex62}
\usepackage{textcomp}
\usepackage{gensymb}

\usepackage[hang,raggedright,tight]{subfigure}
\usepackage{longtable}
\usepackage[figuresright]{rotating}
\usepackage{float}
\usepackage[toc,page]{appendix}

\usepackage{gensymb}
\usepackage{threeparttable}
\usepackage{diagbox}
\usepackage{pbox}
\usepackage{booktabs}
\usepackage{multirow}  
\usepackage{makecell}
\usepackage{natbib}
\usepackage{amsmath,bm}
\usepackage{CJKutf8}
\newcommand{\nxi}{\widehat{\xi}}

\shorttitle{AP test}
\shortauthors{Dong et al.}

\begin{document}
\title{The Most Probable Behaviour of the Dark Energy Equation of State Indicates a Thawing Quintessence Field: Tomographic Alcock–Paczyński Test with Redshift-Space Correlation Function II}
\author[0000-0003-0296-0841]{Fuyu Dong}
\altaffiliation{dfy@ynu.edu.cn}
\affiliation{South-Western Institute for Astronomy Research, Yunnan University, Kunming 650500, People's Republic of China}
\affiliation{Yunnan Key Laboratory of Survey Science, Yunnan University, Kunming, Yunnan 650500, People's Republic of China}
\affiliation{School of Physics, Korea Institute for Advanced Study (KIAS), 85 Hoegiro, Dongdaemun-gu, Seoul, 02455, Republic of Korea}
\author{Changbom Park}
\altaffiliation{cbp@kias.re.kr}
\affiliation{School of Physics, Korea Institute for Advanced Study (KIAS), 85 Hoegiro, Dongdaemun-gu, Seoul, 02455, Republic of Korea}

\begin{abstract} 
We apply an extended Alcock–Paczyński (AP) test to the Sloan Digital Sky Survey data to constrain the dark energy models with the Chevallier–Polarski–Linder (CPL) parametrization of the dark energy equation of state. The extended AP test method uses the full shape of redshift-space two-point correlation funcion(CF) as the standard shape in order to measure the expansion history of the universe. We calibrate the standard shape by using the cosmology-dependent nonlinear evolution of the CF shape in the Multiverse simulations. Further validation of the method and calibration of possible systematics are performed based on mock samples from the Horizon Run 4 simulation. Using the AP test alone, we constrain the flat CDM plus CPL-type dark energy model (flat $w^{\rm CPL}$CDM) to have $\Omega_m=0.290_{-0.031}^{+0.029}$, $w_0=-0.800_{-0.100}^{+0.208}$, and $w_a=-0.238_{-0.972}^{+0.650}$. When combined with other results from the low-redshift universe, such as the Pantheon$+$ supernova compilation and DESI BAO data, the constraint on dark energy becomes $w_0=-0.857_{-0.042}^{+0.051}$, and $w_a=-0.153_{-0.356}^{+0.347}$.
The best-fit $w^{CPL}(z)$ suggests no phantom-divide crossing at $z<0.7$, and the dark energy behaviour is consistent with a thawing quintessence field. It is only when the CMB data are combined with late-time cosmological probes that a phantom-divide crossing at low redshift is favored.

\end{abstract}

\section{Introduction}
The Alcock-Paczyński (AP) test, originally proposed by \cite{1979Natur.281..358A}, exploits the apparent geometrical distortion of intrinsically isotropic structures such as galaxy superclusters to constrain cosmological parameters governing the expansion history of the universe. 
By comparing the observed radial and tangential scales of such objects, one can constrain the product of the Hubble parameter $H(z)$ and angular diameter distance $D_A (z)$ though the AP test.
However, this method is hampered by the non-cosmological shape distortion due to the peculiar velocities of galaxies (redshift-space distortion, hereafter RSD). To overcome this, the extended AP (eAP) test was suggested by \cite{2010ApJ...715L.185P}, and later has been further developed by \cite{2014ApJ...796..137L,2015MNRAS.450..807L,2016ApJ...832..103L} and \cite{2019ApJ...881..146P}. The key idea of the eAP test is to use the redshift evolution of shape rather than shape itself, allowing us to use any shape as far as the shape is conserved.
They found that the shape of two-dimensional redshift-space CF exhibits only very small redshift evolution and the eAP test adopts it as a standard shape.

In \cite{2023ApJ...953...98D}, we have applied this tomographic AP test method to SDSS data, including the SDSS I+II DR7 Main Galaxy Sample, SDSS III BOSS LOWZ and CMASS, and SDSS IV eBOSS LRG samples at $z<0.8$. To extract the shape information of CF in redshift space, we first normalize CF with its volume integral up to the radial separation of $s_{\rm max}$
\begin{equation}
    {\bar \xi}(z)=2\pi\int_0^1d\mu\int_0^{s_{\rm max}}s^2ds \ \xi(s,\mu,z).
\end{equation}
Then to study its angular and radial dependence we compress the information in the normalized CF $\nxi = \xi/{\bar \xi}$ in redshift space into the Legendre multipole moments as in
\begin{equation}
\label{eq:legendre}
     \nxi(s,\mu,z)\approx\sum_{l=0,2,4}\nxi_l(s,z)P_l(\mu),
\end{equation}
where $P_0=1$, $P_2=(3\mu^2-1)/2$ and $P_4=(35\mu^4-30\mu^2+3)/8$. 
An attendant benefit of the normalizing operation is that our results become independent of the overall evolution of clustering amplitude due to gravitational evolution and change of galaxy bias in observational samples.

We have divided the galaxy sample into six nonoverlapping redshift bins, and calculate the likelihood of various cosmological models by comparing the shape evolution of CF between different redshifts:
\begin{equation}
    \Delta\nxi_\ell(s; z_i,z_j)=\nxi_\ell(s; z_i)-\nxi_\ell(s; z_j).
\end{equation}
Using this approach, we constrained the flat $w$CDM model to have $w=-0.892_{-0.050}^{+0.045}$ and $\Omega_m=0.282_{-0.023}^{+0.024}$ from the AP test alone. When combined with the likelihoods from Type I$a$ supernovae (SNe Ia) and SDSS baryon acoustic oscillation (BAO) analyses, the constraints became $w=-0.903_{-0.023}^{+0.023}$ and $\Omega_m=0.285_{-0.014}^{+0.009}$, ruling out the flat $\Lambda$CDM model having $w=-1$  at $4.2\sigma$ level. It is important to note that these tight constraints were resulted even though only low redshift data were used, and are independent of the cosmic microwave background (CMB) observations.  

In addition to the $w$CDM model where the constant $w$ represent an effective value of the dark energy equation of state parameter averaged over a certain redshift interval, \cite{2018ApJ...856...88L} has adopted the Chevallier–Polarski–Linder (CPL) parameterization of dark energy equation of state to examine the potential time evolution of $w$. The CPL DE model describes a time-evolving equation of state using a simple form of $w(a)=w_0+w_a(1-a)$, where $w$ evolves from $w_0+w_a$ at high redshifts to $w_0$ at present.
In combination of the constraint from the eAP test with those derived from
Planck CMB, SDSS-III BAO, JTA SNe Ia, and $H_0$ from HST Cepheids, Li et al. obtained $\Omega_m = 0.301\pm 0.008, w_0 = -1.042\pm 0.067$, and $w_a = -0.07\pm 0.29$ (68.3\% CL). 

Since Li et al.'s work the eAP test has been improved by utilizing the radial shape of correlation function in addition to its angular dependence, resulting in an increase of the constraining power of the test by about 40\% \citep{2019ApJ...881..146P}. Furthermore, a set of cosmological simulations are now available for estimating the cosmology-dependent systematics, which allows further tighter constraints on cosmological parameters \citep{2023ApJ...953...98D}. Most critically, Li et al. has obtained their constraints in combination with the Planck CMB results. As there are many evidence that the cosmology constrained by CMB data tends to be inconsistent with that from low-redshift data \citep{2019ApJ...876...85R,2019MNRAS.484L..64S,2016JCAP...10..019B,2018A&A...609A..72D,2019ApJ...875..145K,2019ApJ...874...32R,2019JCAP...09..006C,2017MNRAS.465.1454H,2013A&A...555A..30B,2015A&A...574A..59D} and the forced combination of statistically mutually excluding results can result in nonsensical results, it is desirable not to combine the constraints of low-redshift probes with CMB results under the currently popular cosmological paradigms.
In this paper we will constrain the CPL parameters $w_0$ and $w_a$ as well as the density parameter $\Omega_m$ using the same SDSS data but without combining the CMB results. Our constraints from the eAP test applied to SDSS will be then combined with those from other low-redshift observations such as  upgraded SN I$a$ data and BAO-only measurements from DESI \citep{2025arXiv250314738D}. 

This paper is organized as follows. \S\ref{sec:CF} introduces the measurement of the two-point CF. \S\ref{sec:data} describes the observational data used in our analysis, while \S\ref{sec:simudata} presents the simulation data. Our constraints on the CPL parameterization are presented in \S\ref{sec:result}. Validation tests performed on the simulation data are discussed in Appendix \ref{sec:appendix}. Finally, we summarize and conclude in \S\ref{sec:conclusion}.

\section{Two Point Correlation Function}
\label{sec:CF}
We use the two-point CF statistic to characterize the clustering of galaxies measured according to the Landy-Szalay estimator \citep{1993ApJ...412...64L}:
\begin{equation}
    \xi(s_\parallel,s_\perp)=\frac{{ DD-2DR+RR}}{{ RR}},
\end{equation}
where $DD, DR$, and $RR$ are the galaxy–galaxy, galaxy–random, and random–random pair counts in redshift space, respectively. $s_\parallel$ is the pair separation parallel to the line of sight (LOS), and $s_\perp$ is the pair separation perpendicular to LOS in comoving coordinates. For a pair of galaxies at positions $\bf{s_1}$ and $\bf{s_2}$ with respect to the observer, we define:
\begin{equation}
    s_\parallel=\frac{\bf{s}\cdot \bf{\bar s}}{|\bf{\bar s}|}, \  s_\perp=\sqrt{\bf{s}\cdot \bf{s}-s_\parallel^2},
\end{equation}
where ${\bf{\bar s}}=({\bf{s_1}}+{\bf{s_2}})/2$, 
${\bf{s}}={\bf{s_2}}-{\bf{s_1}}$.

In this paper, we calculate the two-point CF in polar coordinates, for which the relation can be described as $\mu=s_\parallel/|{\bf s}|=\cos \vartheta$, where $\vartheta$ is the angle between the pair separation $\bf s$ and the line of sight. A pair of galaxies spanning $\Delta z$ in redshift and $\Delta\theta$ in angle have their comoving separations $s_\parallel=[c/H(z)]\Delta z$ and $s_\perp=(1+z)D_A(z)\Delta\theta$
along and across LOS, respectively. When a wrong cosmology is adopted in the conversion from redshift to comoving distance, they are incorrectly changed by the following factors, 
\begin{equation}
\label{EQ:APD}
\alpha_\parallel=\frac{H_{\rm true}(z)}{H_{\rm wrong}(z)}, \ \alpha_\perp=\frac{D_{A,{\rm wrong}}(z)}{D_{A,{\rm true}}(z)}.  
\end{equation}
The AP test exploits the fact that $s_\parallel$ and $s_\perp$ scale differently with the change of cosmology as in Equation \ref{EQ:APD}. This results in apparent changes in shape, $s_\parallel / s_\perp$, or volume, $s_\parallel {s_\perp}^2$.

In our measurement,we choose the Planck 2018 cosmology (TT,TE,EE$+$lowE$+$lensing$+$BAO) to measure DD, DR and RR, and translate these measurements to trial cosmology using the coordinate transforms \citep{2018ApJ...856...88L}:
\begin{equation}
\begin{split}
s_{\rm target}&=s_{\rm fiducial}\sqrt{\alpha^2_\parallel\mu^2_{\rm fiducial}+\alpha_{\perp}^2(1-\mu_{\rm \rm fiducial}^2)},\\
\mu_{\rm target}&=\mu_{\rm fiducial}\frac{\alpha_\parallel}{\sqrt{\alpha^2_\parallel\mu^2_{\rm fiducial}+\alpha_{\perp}^2(1-\mu_{\rm fiducial}^2)}},\\
\end{split}
\end{equation}

\begin{figure}
    \centering
     \subfigure{
     \includegraphics[width=1\linewidth, clip]{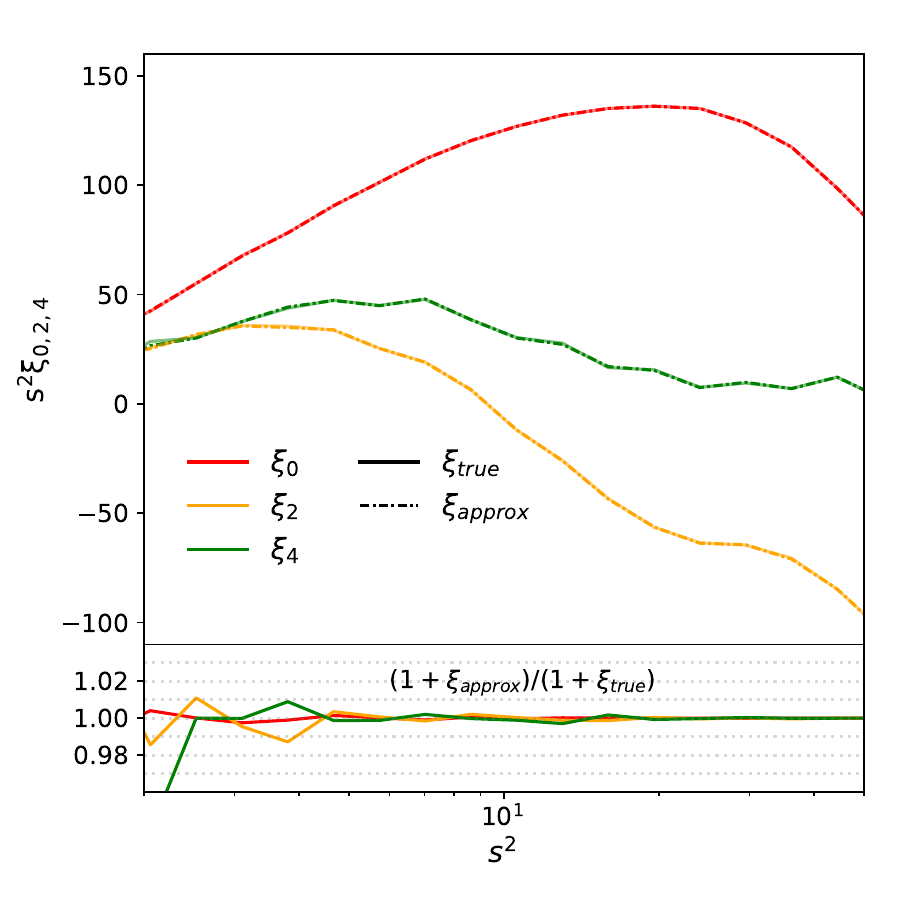}}
   \caption{Comparison of the true and approximate CF. The solid line represents the true value obtained from precise numerical calculations in the cosmology with $\Omega_m=0.26$ and $w=-1.5$, and the dash-dotted line represents the approximation derived from calculation in $\Omega_m=0.31$ and $w=-1$ cosmology and then conversion to $\Omega_m=0.26$ and $w=-1.5$ cosmology using Equation \ref{EQ:APD}. The lower panel shows the difference between them; a constant value of 1 has been added to the CF in this comparison to avoid numerical instabilities near zero.
   } 
    \label{fig:CF}
\end{figure}

We begin by binning galaxy pairs using a high-resolution binning scheme with $\Delta s = 0.1h^{-1}~\text{Mpc}$ and $\Delta\mu = 1/600$. These finely binned pair counts are then aggregated into a coarser binning scheme to compute the DD, DR, and RR correlations in other cosmology. In the coarse scheme, the pair separation $s$ is logarithmically binned from 1 to 60 $h^{-1}\text{Mpc}$ using 20 bins, and the angular coordinate is binned with $\Delta\mu = 1/60$. We also take care of the edge effects by distributing the fine bin into neighboring coarse bins according to the overlapping area fraction. We find that this approximate method is very accurate in transforming the CF measurements between different cosmologies (Figure \ref{fig:CF}).

At last, we choose $s_{\rm max}=60h^{-1}~\text{Mpc}$ in computing the normalization factor. To safely stay away from the scales that are affected by ﬁber collision and Fingers of God, we also exclude the region with $s_\perp <3\, h^{-1}{\rm Mpc}$ of $\nxi$. We then use a $\chi^2$ fitting method to derive the coefficients of the Legendre polynomials $\nxi_{0,2,4}$ \citep{2019ApJ...881..146P}.

\section{Observational Data}
\label{sec:data}
\begin{table}
\centering
\caption{The galaxy samples selected from the SDSS surveys. 
  \label{table: galaxy}}
\centering
\begin{tabular}{c|llll}
\hline
$w$\textbackslash galaxy &redshift &Area[$\rm{deg^2}$] &$N_g$\\
\hline
DR7 &  $0.025\le z<0.163$ &$7.5\times10^3$ &$0.80\times10^5$ \\
LOWZ&  $0.2\le z<0.331$   &$8.6\times10^3$ &$1.57\times10^5$ \\
LOWZ&  $0.331\le z<0.43$  &$8.6\times10^3$ &$1.57\times10^5$ \\
CMASS& $0.45\le z<0.516$  &$9.5\times10^3$ &$2.40\times10^5$\\
CMASS& $0.516\le z<0.576$ &$9.5\times10^3$ &$2.47\times10^5$\\
CMASS& $0.576\le z<0.7$   &$9.5\times10^3$ &$2.44\times10^5$\\
\hline
\end{tabular}
\end{table}

We perform the analysis based on the spectroscopic samples of the SDSS surveys, including the Korea Institute for Advanced Study Value-Added DR7 Main Galaxy Catalog (KIAS-VAGC, \cite{2010JKAS...43..191C}) and the Data Release 12 \citep{2015ApJS..219...12A} of SDSS-III BOSS sample. The BOSS sample consists of two subsamples: `LOWZ' ($0.1\lesssim z\lesssim0.4$) and `CMASS' ($0.4\lesssim z\lesssim0.7$). Following the methodology of \cite{2023ApJ...953...98D}, we conduct separate analyses for each galaxy sample due to their distinct selection criteria and targeting algorithms.

For the DR7 galaxies\footnote{The absolute magnitude cut of $M_r < -21.07 + 5 \log h$ corresponds to a stellar mass threshold of $M_* \sim 10^{11} M_\odot$ for the DR7 sample.}, we impose an $r$-band absolute magnitude cut of $M_r < -21.07 + 5 \log h$ to construct a volume-limited sample covering the redshift range $0.025 < z < 0.163$. For the BOSS survey, we instead apply stellar mass cuts to ensure analysis of only those galaxies with both reliable mass measurements and high sampling completeness in their respective mass ranges.
Specifically, we set: $M_\star > 10^{11.1} M_\odot$ for LOWZ galaxies, $M_\star > 10^{11} M_\odot$ for CMASS galaxies.
The slightly lower mass threshold for CMASS accounts for its systematically less massive population compared to LOWZ. As summarized in Table. \ref{table: galaxy}, we further divide the galaxies in all the catalogs into a toal of six subsamples.

The stellar masses of galaxies in our analysis are drawn from multiple sources with appropriate calibrations. For the DR7 sample, we utilize the publicly available MPA-JHU catalog values as our reference stellar masses\footnote{\url{https://wwwmpa.mpa-garching.mpg.de/SDSS/DR7/}}. For the BOSS galaxies, we adopt the stellar mass estimates from \cite{10.1111/j.1365-2966.2011.20306.x}, who derived masses through principal component analysis (PCA) of galaxy spectra using:
\begin{itemize}
\item The \cite{2003MNRAS.344.1000B} stellar population synthesis (SPS) models
\item A \cite{2001MNRAS.322..231K} initial mass function (IMF)
\item The dust attenuation model of \cite{2000ApJ...539..718C}
\end{itemize}
Following \cite{2018ApJ...858...30G}, we apply a systematic offset of $-0.155$ dex to these BOSS masses to ensure consistency with the literature. This correction accounts for:
\begin{itemize}
\item A $-0.105$ dex overestimation inherent to the PCA method relative to MPA-JHU masses
\item An additional $-0.05 dex$ adjustment for IMF conversion from \cite{2001MNRAS.322..231K} to Chabrier (2003)
\end{itemize}

The random catalogs for the LOWZ and CMASS
samples, which incorporate the angular selection functions, are provided alongside the observational data. We sample random points from these catalogs while preserving the redshift distribution of our galaxy samples.
For BOSS target, we apply three distinct weighting schemes to mitigate systematic effects:
\begin{itemize}
\item WEIGHT\_CP: Corrects for fiber collision effects (close pairs)
\item WEIGHT\_NOZ: Accounts for redshift measurement failures
\item WEIGHT\_SYSTOT: Compensates for imaging systematics
\end{itemize}
The combined weight for each galaxy and random point in our analysis is calculated as 
\begin{equation}
w_{\rm total} = (w_{\rm cp} + w_{\rm noz} - 1) \times w_{\rm sys}
\end{equation}

\section{Simulation Data}
\label{sec:simudata}
In addition to the Alcock-Paczyński (AP) effect, two distinct physical effects contribute to redshift-dependent anisotropy in galaxy clustering patterns: (1) redshift-space distortions (RSDs) and (2) survey selection effects.  RSDs - arising from galaxy peculiar velocities - constitute the dominant source of systematic uncertainty in our analysis: it modifies the CF shape across both linear and nonlinear scales, thus introducing scale-dependent anisotropies in clustering patterns. Furthermore, survey selection functions (both angular and radial) introduce additional systematic effects through:
\begin{itemize}
\item Inhomogeneous sky coverage
\item Redshift-dependent completeness variation
\item Galaxy bias evolution across samples
\end{itemize}
To address these challenges, we employ two state-of-the-art simulation suites\citep{2023ApJ...953...98D}:
\begin{itemize}
\item The high-resolution Horizon Run 4 (HR4) N-body simulation
\item An ensemble of Multiverse simulations 
\end{itemize}

\subsection{Horizon Run 4}
\label{sec:hr4}
The HR4 simulation has a box size of 3150 $h^{-1}{\rm Mpc}$, and employs $N=6300^3$ dark matter particles \citep{2015JKAS...48..213K}. It adopts the WMAP 5-year cosmology, with $(\Omega_{\Lambda},\Omega_m,\Omega_b,h,\sigma_8,n_s)=(0.74,0.26,0.044,0.72$, $0.794,0.96)$. We take the most-bound dark matter particle of each halo as the galaxy \citep{2016ApJ...823..103H}, and produce mock `galaxies' by using the modified merger time scale model of \cite{Jiang_2008}. To generate realistic mock catalogs, we transform the simulated galaxy positions from real space to redshift space that match each survey's geometry. The redshift-space conversion incorporates both cosmological expansion and peculiar velocities through the relation:
\begin{equation}
\label{eq:zrsd}
z_{\rm obs} = z_{\rm cos} + \frac{v_{\rm pec}}{c}(1+z_{\rm cos})
\end{equation}
where $z_{\rm cos}$ is the cosmological redshift from Hubble flow and $v_{\rm pec}$ represents the line-of-sight peculiar velocity component. For each observed sample, we select galaxies from the simulation snapshot closest in cosmic time to the target redshift range\footnote{The HR4 simulation provides snapshots at z = 0, 0.05, 0.1, 0.15, 0.2, 0.3, 0.4, 0.5, 0.6, 0.7, and 1.}. Using this methodology, we generate 8(6) independent mock realizations for each BOSS sample.  These mock catalogs serve two main purposes in our analysis:
\begin{itemize}
\item To quantify systematic errors in the observed correlation function
\item To validate the analysis pipeline
\end{itemize}

To reproduce the observed number densities of BOSS galaxies in our mock catalogs, we employ stellar mass as our primary mass proxy. We implement a subhalo abundance-matching approach \citep{2004MNRAS.353..189V} to assign stellar masses to mock galaxies, ensuring consistency with the observed distributions. For the DR7 sample, we create volume-limited mocks in a similar way within the redshift range of $0.025<z<0.163$. We have created 108 mock DR7 survey catalogs in total extracted from the HR4 simulation. We select galaxies with $r$-band absolute magnitudes brighter than $-21.07$ after luminosity evolution correction is applied (see \cite{2007ApJ...658..884C,2010JKAS...43..191C} for details).

\subsection{Multiverse Simulations}
\label{sec:multiverse}
The Multiverse simulations consist of a suite of N-body simulations modeling cold dark matter (CDM) universes with either constant or time-varying dark energy equations of state. These simulations employ a periodic comoving box of 1024 $\rm{Mpc}/h$ containing $2048^3$ dark matter particles. Furthermore, all simulations employ identical random number sequences to generate their initial conditions, ensuring exactly matched cosmic variance across simulations. This design isolates differences arising purely from cosmological parameter variations. Each model’s power spectrum is normalized to produce the same root-mean-square matter fluctuation amplitude ($\sigma_8=0.794$) at $z=0$.

These simulations enable robust comparisons between cosmologies, even on non-linear scales, free from cosmic variance effects.  For this study, we analyze ten Multiverse simulations with constant dark energy equations of state (Table 2 in \cite{2023ApJ...953...98D}). All adopt a flat geometry and share the same remaining cosmological parameters ($\Omega_b$, $h$, $\sigma_8$, $n_s$) with the HR4 simulation.

For the full-sky DR7 mocks, we position an observer at the center of the simulation box and omit angular masking to minimize statistical uncertainties in the correlation function measurements. For LOWZ and CMASS samples at higher redshifts, we generate corresponding mocks within a cubic box rather than using spherical slices, as the limited box size of Multiverse simulations makes spherical slice mocks unfeasible. To include the RSD effects within a cubic box,  we treat the (comoving) $x_3$ axis of the simulation box as the LOS direction and use the following equation to shift the LOS coordinate of a mock galaxy at $z_i$ from real to redshift space:
\begin{equation}
\widetilde{x}_3 = x_3 + v_3\frac{1+z_i}{H(z_i)},
\end{equation}
where $\widetilde{x}_3$ denotes the shifted $x_3$-coordinate, $v_3$ is the proper peculiar velocity of the galaxy in real space along $x_3$ direction, and $H(z)$ the Hubble parameter. We also generate two other catalogs by choosing $x_1$ and $x_2$ as the LOS direction respectively. As the last step, we average the CF measured from three mocks to reduce the statistical uncertainty.

As the Multiverse simulations having about 1 $h^{-1}$Gpc size are too small to be used to generate mock survey samples, they are used only to capture the CF shape difference between different cosmological models.
To estimate the full amount of the nonlinear systematics in the redshift evolution of CF, we will use the CFs of Multiverse simulations in combination with the CF measured from the HR4 mocks. We also account for the effects of mass-selection incompleteness  by using an average mass selection function for each redshift slice at $z_i$.

\subsection{Intrinsic redshift evolution of the shape of CF}
\label{sec:nxi-sys}
We make accurate estimation of the systematic shape evolution of CF in HR4 as the influence of cosmology and nonlinear evolution on the shape of CF becomes non-negligible relative to geometric distortions when high-precision in cosmological parameters is required \citep{2019ApJ...881..146P}. To account for this dependence, we used the Multiverse simulations to estimate the systematic effects on the normalized CF shape difference $\Delta {\hat \xi}(s,\mu,z_i,z_j)$ between two redshifts in each multiverse simulation cosmology $M$
by considering only deviations from the fiducial model:
\begin{equation}
\begin{split}
\Delta\nxi^{\rm{sys}}(\Omega_m^{\rm M}, w^{\rm M})=&\Delta\nxi^{{\rm M}}(\Omega_m^{\rm M},w^{\rm M})-\Delta\nxi^{{\rm M}}(\Omega^{{\rm M5}}_{m},w^{{\rm M5}}) +\\
&\Delta\nxi^{{\rm HR4}}(\Omega^{{\rm HR4}}_{m},w^{{\rm HR4}}),
\end{split}
\end{equation}
where $\Omega^{{\rm M5}}_{m}=\Omega^{{\rm HR4}}_m =0.26, w^{{\rm M5}}=w^{{\rm HR4}}=-1$, which are the parameters of the fiducial model or HR4 cosmology.

We model the $\mu$- and $s$-dependence of $\Delta {\hat \xi}(s,\mu)$ by expanding it into Legendre multipole moments as in Equation (2) and approximating their coefficients into 
a second-order polynomial in $\log s$:
\begin{equation}
    s^2\Delta\nxi_{\ell}^{\rm sys}\approx a_{\ell}^{\rm sys}+b_{\ell}^{\rm sys}{\rm log}~s+c_{\ell}^{\rm sys}({\rm log} ~s)^2.
\end{equation}
To get the systematic shape evolution for any $w$CDM cosmology, 
we interpolate or extrapolate the coefficients from those of $\Delta\nxi^{\rm{sys}}(\Omega_m^{\rm M}, w^{\rm M})$ to any $w$CDM cosmology using a third-order polynomial \citep{2023ApJ...953...98D}.

To further estimate the intrinsic redshift evolution across different CPL model parameters, we approximate the effective equation of state parameter $w_{\rm eff}$ using the following expression:
\begin{equation}
    w_{\rm eff}=\frac{\int_0^{z_{\rm max}}\Omega_{\rm DE}(z)w(z)\frac{dV}{dz}dz}{\int_0^{z_{\rm max}}\Omega_{\rm DE}(z)\frac{dV}{dz}dz},
\end{equation}
thereby incorporating nonlinear corrections derived from the procedure above.

\section{Result}
\label{sec:result}
\subsection{Likelihood Analysis}
\label{sec:likelihood}

We choose a reference redshift, $z_{\rm ref}$, and measure the CF shape difference between $z_i$ and $z_{\rm ref}$
\begin{equation}
\Delta\nxi(z_i,z_{\rm ref})=\nxi(z_i)-\nxi(z_{\rm ref})-\Delta\nxi^{\rm{sys}}(z_i,z_{\rm ref}), 
\end{equation}
where $z_i\neq z_{\rm ref}$ and $\Delta\nxi^{\rm{sys}}$ is the correction taking care of the intrinsic shape change due to gravitational evolution, survey sample variation, ﬁber-collision effect, and various other observational effects \citep{2023ApJ...953...98D}. $\Delta\nxi^{\rm{sys}}$ is obtained as described in the previous section. For the likelihood estimation below, we use the data points of $\Delta\nxi$ within the radius range of $s = 6–15\, \rm{Mpc/h}$ \citep{2019ApJ...881..146P,2023ApJ...953...98D}.

We perform a joint analysis taking into account the correlations between different $z_i$ bins for a fixed $z_{\rm ref}$ 
The likelihood for any adopted cosmology and the reference redshift $z_{\rm ref}$ chosen is given by $\mathcal{L}_{z_{\rm ref}}={\rm exp}(-\chi^2_{z_{\rm ref}}/2)$, where
\begin{equation}
\label{eq:chi2}
    \chi^2_{z_{\rm ref}}(\Omega_m,w)=\sum_{\ell=0,2,4}\sum_{\alpha,\beta} P_{\ell}^{z_{\rm ref}}(s_\alpha)\cdot (C^{\ell}_{\alpha\beta})^{-1} \cdot P_{\ell}^{z_{\rm ref}}(s_\beta).
\end{equation}
Here $P_{\ell}^{z_{\rm ref}}=(...,\Delta\nxi_{\ell}(z_i,z_{\rm ref}),...,\Delta\nxi_{\ell}(z_j,z_{\rm ref}),...)$ ($z_i \neq z_{\rm ref}\neq z_j$, $i<j$), $s_\alpha$ is the $\alpha$-th pair separation bin radius, and $C^{\ell}_{\alpha\beta}$ the covariance matrix of the ${\ell}$-th moment in the Legendre polynomial expansion. One can take any redshift as the reference $z_{\rm ref}$. The likelihood estimates from different $z_{\rm ref}$ are found consistent after incorporating cross-correlations between redshifts. The final result is taken as the average over the six reference redshifts: $\mathcal{L}=\Sigma_{i=1}^n\mathcal{L}_{z_i}$/n, where $n=6$ and $\mathcal{L}_i$ is the likelihood when $z_i$ is chosen as the reference redshift.

To assess the covariance of CF we need a large number of mock samples. We used the MultiDark PATCHY mock catalogs \citep{2016MNRAS.456.4156K} to make mocks for the BOSS DR12 data. Both PATCHY and EZmock generate the dark matter density field using the Zel'dovich approximation \citep{1970A&A.....5...84Z}. The PATCHY mocks have been calibrated using a reference simulation to obtain a detailed galaxy bias evolution spanning the redshift range from 0.15 to 0.75. For the DR7 sample, we used 108 mocks from the HR4 $N$-body simulation to assess the covariance.

\begin{figure*}
    \centering
     \subfigure{
     \includegraphics[width=0.7\linewidth, clip]{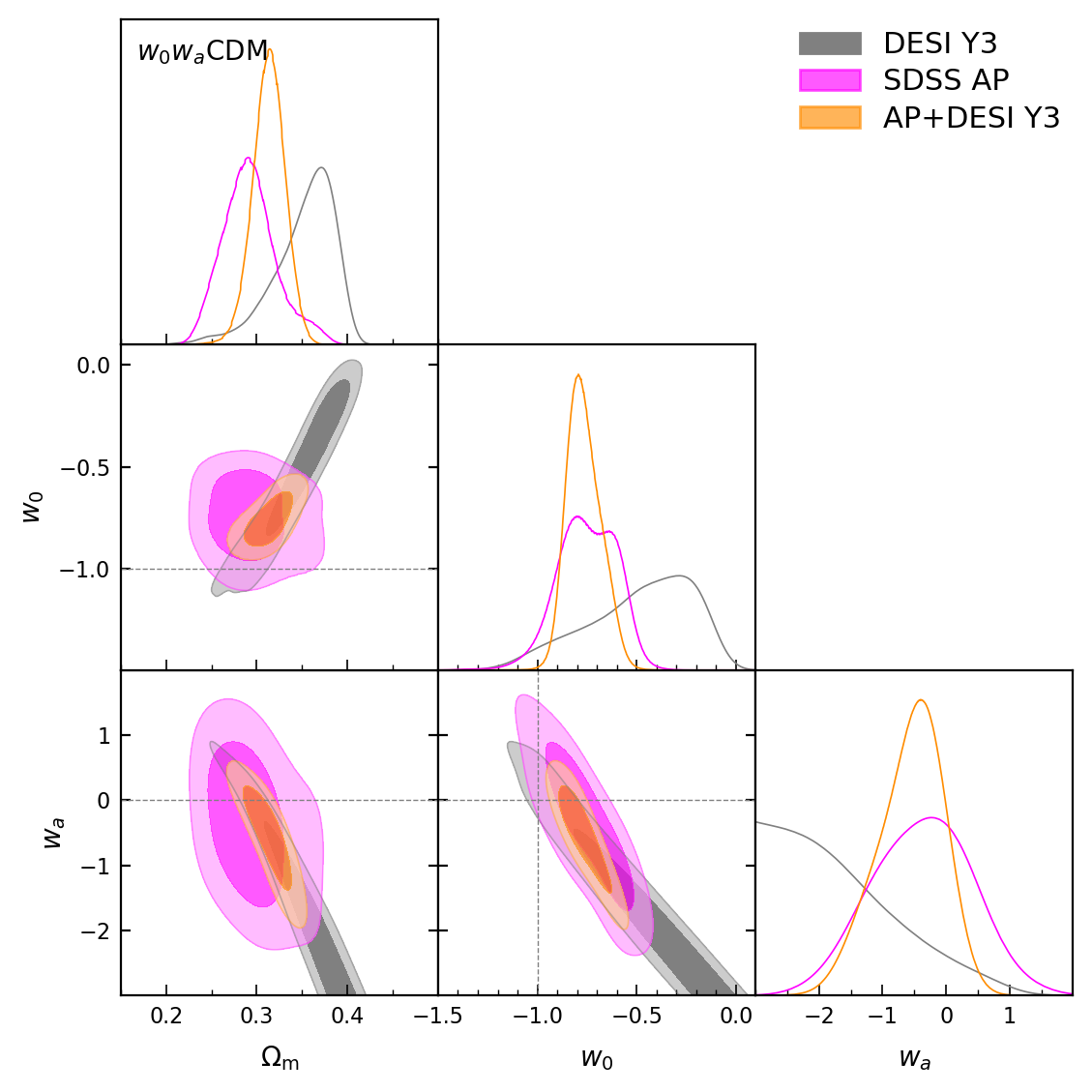}}
   \caption{Constraints on the flat $w^{\rm CPL}$CDM model from SDSS AP-only (magenta), DESI-Y3 BAO-only (DR2, grey), and SDSS AP+DESI BAO (orange) analyses. The sub-panels show the marginalized probability distribution function of each cosmological parameter. The likelihood contours from the BAO analysis and AP test exhibit some overlapping distributions in the parameter space. However, the AP test shows a good consistency with $w_a=0$ while BAO favors $w_a<0$.
   }  
    \label{fig:result_APBAO}
\end{figure*}

\begin{figure*}
    \centering
     \subfigure{
     \includegraphics[width=0.7\linewidth, clip]{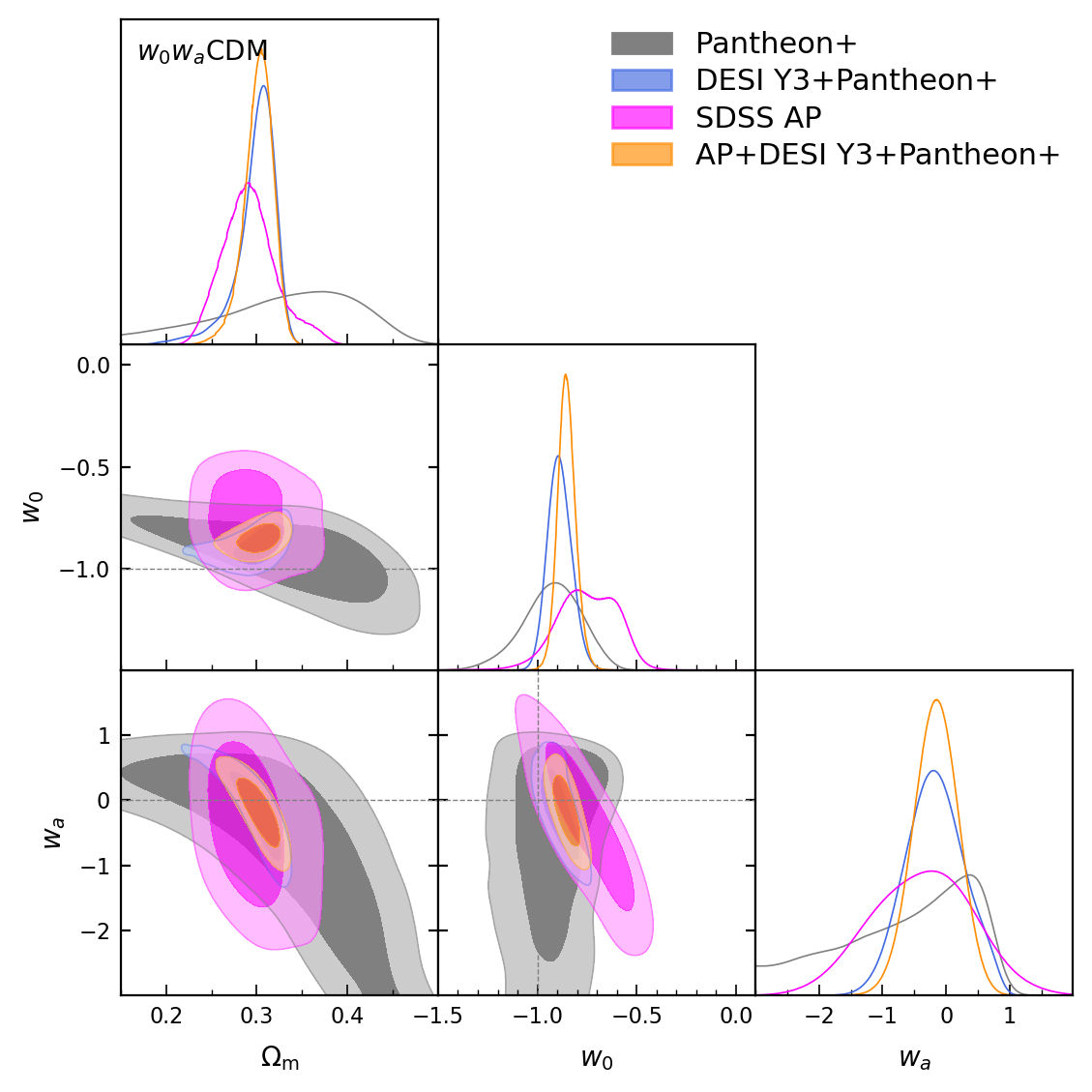}}
   \caption{Similar to Figure \ref{fig:result_APBAO}, but Pantheon$+$ SNe I$a$ data (grey) is included for joint-constraints. The orange contours are from combining the constraints of three low-redshift probes of SDSS AP, DESI BAO, and Pantheon$+$ SN I$a$. The probability distribution of $w_0$ and $w_a$ from the joint analysis indicates that the dark energy equation of state is most likely to evolve slowly with no phantom-divide crossing up to $z=0.7$.
   }  
    \label{fig:result_APSNeBAO}
\end{figure*}

\subsection{Constraint on the flat $w^{\rm CPL}$CDM models}
In this section, we will constrain the flat $w^{\rm CPL}$CDM models.
In these models the AP test is applied to constrain three free parameters, 
$\theta=\{\Omega_m,w_0,w_a\}$.
We apply our method to the six redshift samples of the SDSS survey, following the likelihood analysis described in \S\ref{sec:likelihood}. We explore the three-dimensional parameter space spanned by $\Omega_m$,$w_0$ and $w_a$ using flat priors within the ranges [0.1,0.5], [$ -1.5,0$] and [$-3$,2], with grid intervals of $\Delta \Omega_m=0.01$, $\Delta w_0=0.03$ and $\Delta w_a=0.1$, respectively. At each grid point in the parameter space ($\Omega_m$,$w_0$,$w_a$), we compute the CF of the galaxies based on the corresponding cosmology. 

The magenta contours in Figure \ref{fig:result_APBAO} are the results of our SDSS AP test, and  the best-fit values and uncertainty limits are given in Table \ref{Table-result}. The constraints on the parameters from our AP test alone are
$\Omega_m=0.290_{-0.031}^{+0.029}, w_0=-0.800_{-0.100}^{+0.208},w_a=-0.238_{-0.972}^{+0.650}$.
An anti-correlation is observed in the $w_0$–$w_a$ plane, which has a straightforward physical interpretation: in the CPL parameterization
of dark energy equation of state
$w$ evolves from $w_0 + w_a$ to $w_0$. When $w_0$ and $w_a$ are chosen to be anti-correlated, the average value of $w$ is roughly unchanged and the expansion history is similar. 
$w_0$ is more tightly constrained than $w_a$
because the latter is modulated by a factor of $1-a=z/(1 + z)$ in the expression for $w(z)$ which diminishes its influence, especially at low redshifts. 

Comparisons of the AP results with those of DESI BAO \citep{2025PhRvD.112h3515A} are also shown in the same figure (grey) and table. The constraints on cosmological parameters from BAO alone are relatively weak, and there is no meaningful constraint on $w_a$ in particular. A fatal problem with the DESI BAO result is that a flat prior of [$-3$, 2] is applied to $w_a$. This not only affects the pdf of $w_a$, but also the pdfs of $\Omega_m$ and $w_0$.

While the two sets of constraints from AP and BAO show an overlap within the 68\% confidence regions across the marginalized 2D planes of $\{\Omega_m–w_0, \Omega_m–w_a, w_0–w_a\}$ under the DESI BAO's limited flat prior, the AP results favor relatively smaller $\Omega_m$ and more negative $w_0$. 
Most importantly, the AP test indicates that $w_a$ is consistent with zero with its most probable value being slightly negative, suggesting that the dark-energy equation of state may decrease only slowly toward higher redshifts.

To complement our analysis and investigate DE further, we extend the analysis by combining the constraint from the Pantheon$+$ SNe I$a$ sample. Compared with the original `Pantheon sample' which consisted of 1,048 spectroscopically confirmed SNe I$a$ from PanStarrs \citep{10.1117/12.859188}, supplemented by low-redshift SNe I$a$ \citep{1999AJ....117..707R,2006AJ....131..527J,2009ApJ...700.1097H,2010AJ....139..519C,2010AJ....139..120F,2011AJ....142..156S}, as well as samples from SDSS, SNLS, and HST \citep{2012ApJ...746...85S,2007ApJ...659...98R,2014AJ....148...13R,2014ApJ...783...28G,2018ApJ...853..126R}-the Pantheon$+$ SNe I$a$ dataset incorporates six additional samples \citep{2022ApJ...938..113S}. It comprises 1,550 spectroscopically-confirmed SN Ia in the redshift range $0.001 < z < 2.26$, with the most substantial increase in supernova counts occurring at low redshifts.

We use the public likelihood from \cite{2022ApJ...938..110B}, incorporating the full statistical and systematic covariance. The parameter constraints from AP (magenta), BAO, and Pantheon$+$ (grey) are shown in Figure  \ref{fig:result_APSNeBAO}. The combined measurement of the time-evolution parameter $w_a$ is $-0.153_{-0.356}^{+0.347}$, which remains consistent with a non-evolving or weakly-evolving dark energy equation of state. It should be noted that the most probable $w^{CPL}(z)$ indicates no phantom-divide crossing of dark energy within the explored redshift range, namely up to $z=0.7$, and that the dark energy behaviour is consistent with a thawing quintessence field. 
Interestingly, combining only the BAO and SN data (blue contours) yields 
$w_a = -0.188_{-0.451}^{+0.493}$, a result consistent with that obtained from all three low-redshift probes combined. This statistical agreement between the AP and BAO+SN constraints is encouraging and justifies combining them.

Consistent cosmological constraints are obtained from the AP(joint)+BAO+SN dataset when accounting for mode correlations: $\Omega_m=0.299_{-0.019}^{+0.011}, w_0 = -0.880_{-0.046}^{+0.048}, w_a=-0.096_{-0.379}^{+0.340}$. The AP(joint) is derived from a joint analysis of the three radial modes expanded in Legendre polynomials ($P_0$, $P_2$, $P_4$), employing the full covariance matrix that accounts for their cross-correlations ($P_0-P_2$, $P_2-P_4$, $P_4-P_0$)\footnote{The joint analysis yields overall consistent constraints while shifting the parameter $w_a$ from $-0.238_{-0.972}^{+0.650}$ to $-0.728_{-0.750}^{+0.889}$ (but with more positive value of $w_0 = -0.643_{-0.243}^{+0.038}$)}.

\begin{table*}
\centering
\caption{
Marginalized best-fit values, $\Omega_m$, $w_0$ and $w_a$, average values, $\langle\Omega_m\rangle$, and $\langle w_0\rangle$ and $\langle w_a\rangle$, and their 68$\%$ confidence limits
estimated from various cosmological probes and their combined likelihood analyses.
Our main result is from combination of SDSS AP $+$ DESI BAO $+$ Pantheon$+$ SN.
  \label{Table-result}}
\begin{minipage}{180mm}
\centering
\begin{tabular}{c|cc|cc|cc}
\hline
\hline
probes&$\Omega_m$(best) & $\langle\Omega_m\rangle$ &$w_0$(best) &$\langle w_0 \rangle$ & $w_a$(best) & $\langle w_a \rangle$\\
\hline
BAO            & $0.371_{-0.053}^{+0.013}$&$0.352\pm0.035$ &$-0.285_{-0.490}^{+0.078}$&$-0.480\pm0.268$ &$<-1.32$ &-\\
SN             & $0.370_{-0.149}^{+0.039}$&$0.318\pm0.095$ &$-0.910_{-0.158}^{+0.134}$&$-0.925\pm0.147$ &$0.374_{-2.201}^{+0.087}$ &$-0.606\pm1.032$\\
AP             & $0.290_{-0.031}^{+0.029}$&$0.292\pm0.030$ &$-0.800_{-0.100}^{+0.208}$&$-0.749\pm0.148$ &$-0.238_{-0.972}^{+0.650}$ &$-0.380\pm0.802$\\
BAO+SN         & $0.308_{-0.027}^{+0.011}$&$0.299\pm0.024$ &$-0.895_{-0.052}^{+0.068}$&$-0.887\pm0.060$ &$-0.188_{-0.451}^{+0.493}$&$-0.172\pm0.461$\\
AP+BAO         & $0.314_{-0.014}^{+0.016}$&$0.313\pm0.018$ &$-0.794_{-0.053}^{+0.121}$& $-0.765\pm0.085$&$-0.393_{-0.734}^{+0.353}$&$-0.554\pm0.525$\\
AP+SN          & $0.295_{-0.025}^{+0.025}$&$0.296\pm0.029$ &$-0.861_{-0.079}^{+0.069}$&$-0.867\pm0.077$ &$0.124_{-0.515}^{+0.405}$ &$0.064\pm0.460$\\
AP+BAO+SN      & $0.305_{-0.015}^{+0.015}$&$0.302\pm0.016$ &$-0.857_{-0.042}^{+0.051}$&$-0.852\pm0.045$ &$-0.153_{-0.356}^{+0.347}$ &$-0.161\pm0.348$\\
\hline
\end{tabular}
\end{minipage}
\end{table*}

 In the Appendix, we validate our method using mock data and assess potential systematic uncertainties in the parameter constraints, which are incorporated into the calibration of our method.

\begin{figure}
    \centering
     \subfigure{
     \includegraphics[width=0.95\linewidth, clip]{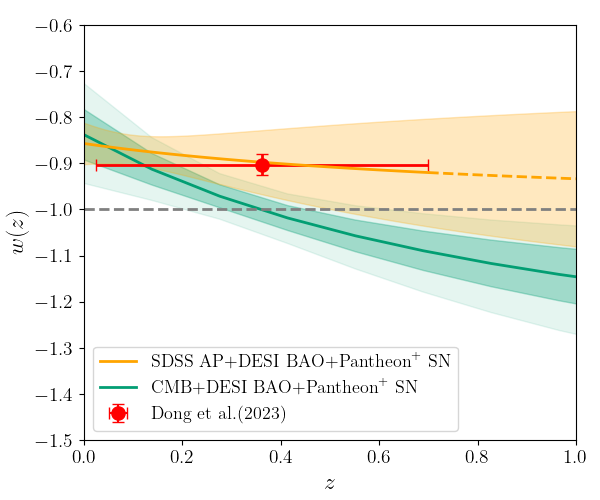}}
   \caption{The most probable evolution of the dark energy equation of state parameter $w(z)$ within the CPL parametrization, obtained from combining SDSS AP with DESI BAO and Pantheon$+$ SN data (orange line).
   The light orange contour represents the $68\%$ confidence region. The red dot denotes the Dong et al. (2023)'s measurement of $w_{\rm eff}$ within the flat $w$CDM paradigm derived from the same data. The horizontal error bar represents the redshift range of the observational samples. For comparison, 
   the green line shows the corresponding constraints derived from CMB + DESI BAO + Pantheon$+$ SN\citep{2025PhRvD.112h3515A}.
   } 
    \label{fig:wz}
\end{figure}
\begin{figure}
    \centering
     \subfigure{
     \includegraphics[width=0.95\linewidth, clip]{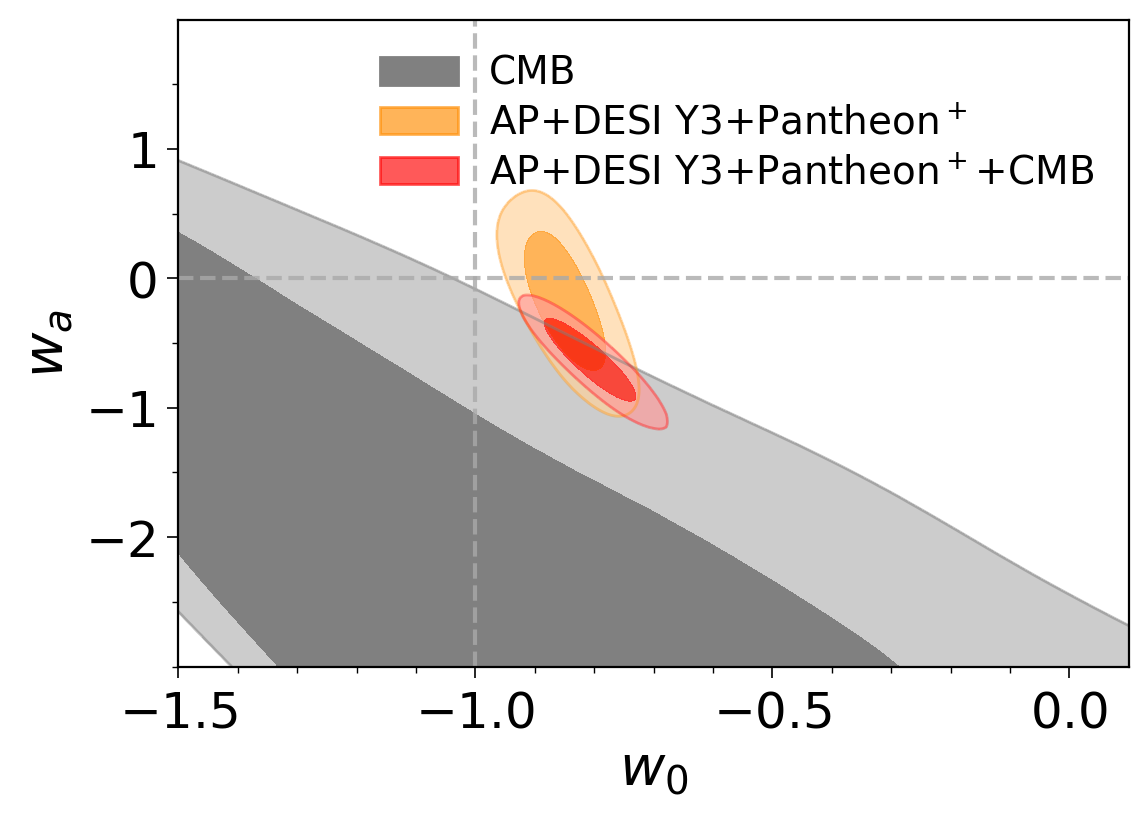}}
   \caption{Results for the posterior distributions of $w_0$ and $w_a$, derived from the combined constraints of SDSS AP, DESI BAO and Pantheon+ SN data on the flat $w^{\rm CPL}$CDM model, excluding (orange) and including CMB (red). The CMB has relatively much weaker constraining power and the favored region in the $w_0$-$w_a$ plane is greatly shifted from those of low-redshift probes. The inclusion of CMB data significantly pulls down the cosmological constraint away from $w_a=0$, driving it to a phantom-divide crossing scenario.
   } 
    \label{fig:w0wa2d}
\end{figure}

\section{Summary and Discussion}
\label{sec:conclusion}
We perform an extended AP test adopting the normalized redshift-space two-point CF of galaxies as the standard shape to constrain the flat $w^{\rm CPL}$CDM model. 
The test is applied to the SDSS spectroscopic galaxy redshift survey sample, which contains about $1.25\times 10^6$ galaxies in the redshift range 
$0.025 < z < 0.7$. 
The results of the current analysis are quite consistent with our previous results (Dong et al. 2023) for the flat $w$CDM models, $\Omega_m = 0.282^{+0.024}_{-0.023}$ and $w=-0.892^{+0.045}_{-0.050}$ (see the $w_a=0$ cuts of the magenta contours in Figure 2).
The most notable outcome of the current AP-test analysis constraining flat $w^{\rm CPL}$CDM models is that $w_a$ is consistent with zero, with its most probable value being slightly negative, supporting a non-evolving or only weakly evolving dark-energy equation of state
We note that the combination of the DESI BAO and Pantheon$+$ SNe I$a$ results also yields a statistically same conclusion.

Since the standard-shape analysis with the AP test, the standard-ruler analysis with BAO scale measurements, and the standard-candle analysis with SN luminosity distances are statistically consistent with one another — and more importantly since they lead to the same conclusion regarding the physical properties of dark energy — we combine the three probes to obtain a joint set of constraints
$\Omega_m=0.305_{-0.015}^{+0.015}, w_0=-0.857_{-0.042}^{+0.051}, w_a=-0.153_{-0.356}^{+0.347}$.

In the $\Omega_m-w_a$ and $w_0-w_a$ planes, the AP test exhibits degeneracy directions aligned with those of BAO, though the confidence region of BAO is relatively much more extended.  
While SN constraints display a degeneracy direction distinct from both BAO and AP, yet show good positional agreement with the AP results. The addition of the AP test significantly improves the constraining power compared to using BAO or SN alone. Owing to the good consistency between the AP and the BAO+SN constraints, the incremental improvement from adding the AP test is modest.
Nonetheless, their consistency highlights the robustness of the cosmological results.

In Figure \ref{fig:wz} we compare our result with other recent measurements of $w$ as a function of redshift. First, the red point shows the average $w$ over $0.025<z<0.7$ measured by Dong et al. (2023), who adopted the flat $w$CDM cosmology and combined constraints from the SDSS AP test, SDSS BAO, and Pantheon SN data. This measurement ruled out the flat $\Lambda$CDM model at the $4.2\sigma$ level.
The orange line shows the main result of this work: the constraints on $w(z)$ under the CPL parametrization, derived from combining our AP test with DESI BAO and Pantheon$+$ data. 
The best-fit $w^{\rm CPL}(z)$ remains above $-1$ and approaches $-1$ only slowly as redshift increases indicating no phantom-divide crossing within the redshift range explored ($z<0.7$). This behaviour of the equation of state suggests that dark energy might be a thawing quintessence field.
By contrast, the best-fit CPL model constrained by CMB + DESI BAO + Pantheon$+$ shown as the green line favors a phantom-divide crossing at $z\approx 0.36$.

The phantom-divide crossing in $w(z)$ inferred from the combination of CMB + DESI BAO + Pantheon+ arises primarily from the inclusion of the CMB data. In the absence of the CMB constraint, the joint DESI BAO and Pantheon+ SN data favor no phantom-divide crossing. Figure~\ref{fig:w0wa2d} illustrates that, although the CMB has relatively weak standalone constraining power in the $w_0$-$w_a$ plane, it has a disproportionate impact on the inferred value of 
$w_a$. When the CMB constraint (grey) is combined with the joint late-time probes (orange), 
$w_a$ is driven significantly toward more negative values. Consequently, the best-fit CPL parameters shift from $(w_0, w_a)=(-0.86,-0.15)$ to $(-0.81, -0.62)$ (red), thereby inducing a phantom-divide crossing in $w(z)$.


These findings demonstrates the considerable potential of the AP test for constraining the expansion history of the universe and thus the dark energy equation of state. However, due to limitations in the galaxy sample—such as finite spatial coverage and limited number density of galaxies—prevent high-precision constraints on $w_a$ using the AP test alone ($\sigma_{w_a}\sim0.8$). To address these limitations, our next step involves the analysis of DESI Data Release 2 (DR2) and analyses of different dark energy models. 
This dataset provides a much increased galaxy density and a wider angular coverage over an increased redshift range.
These improvements will substantially reduce statistical uncertainties in the AP test and mitigate cosmic variance. 
In particular, DESI DR2 reaches higher redshifts than SDSS, so the constraints from the AP test are expected to be tighter as the test uses an extended range of the expansion history of the universe.

\section*{Acknowledgments}

FD acknowledges the ﬁnancial support from the National Natural Science Foundation of China, grant No.12303003, and Yunnan Key Laboratory of Survey Science, grant No.202449CE340002.
CBP is supported by the KIAS Individual Grant PG016904 at the Korea Institute for Advanced Study (KIAS). CBP is also supported by the National Research Foundation of Korea (NRF) grant funded by
the Korean government (MSIT; RS-2024-00360385).

\typeout{}
\bibliography{ap}
\appendix

\section{Method Validation and systematics calibration}
\label{sec:appendix}
Potential sources of systematic bias in our methodology include: (i) inaccuracies in the nonlinear evolution corrections, (ii) inaccuracies in interpolating the 2pt CF from the baseline cosmology to other cosmological models, and (iii) inherent limitations of the extended AP test itself. In this work we use realistic mock survey samples that incorporate all the important aspects of observation such as survey angular and radial selection functions and galaxy mass sampling rate variation in order to test our method. We will calculate the likelihood of various cosmological models when the true cosmology is that of HR4. The mock survey samples drawn from HR4 are used in this calculation.

\begin{figure*}[!htb]
    \centering
     \subfigure{
     \includegraphics[width=0.6\linewidth, clip]{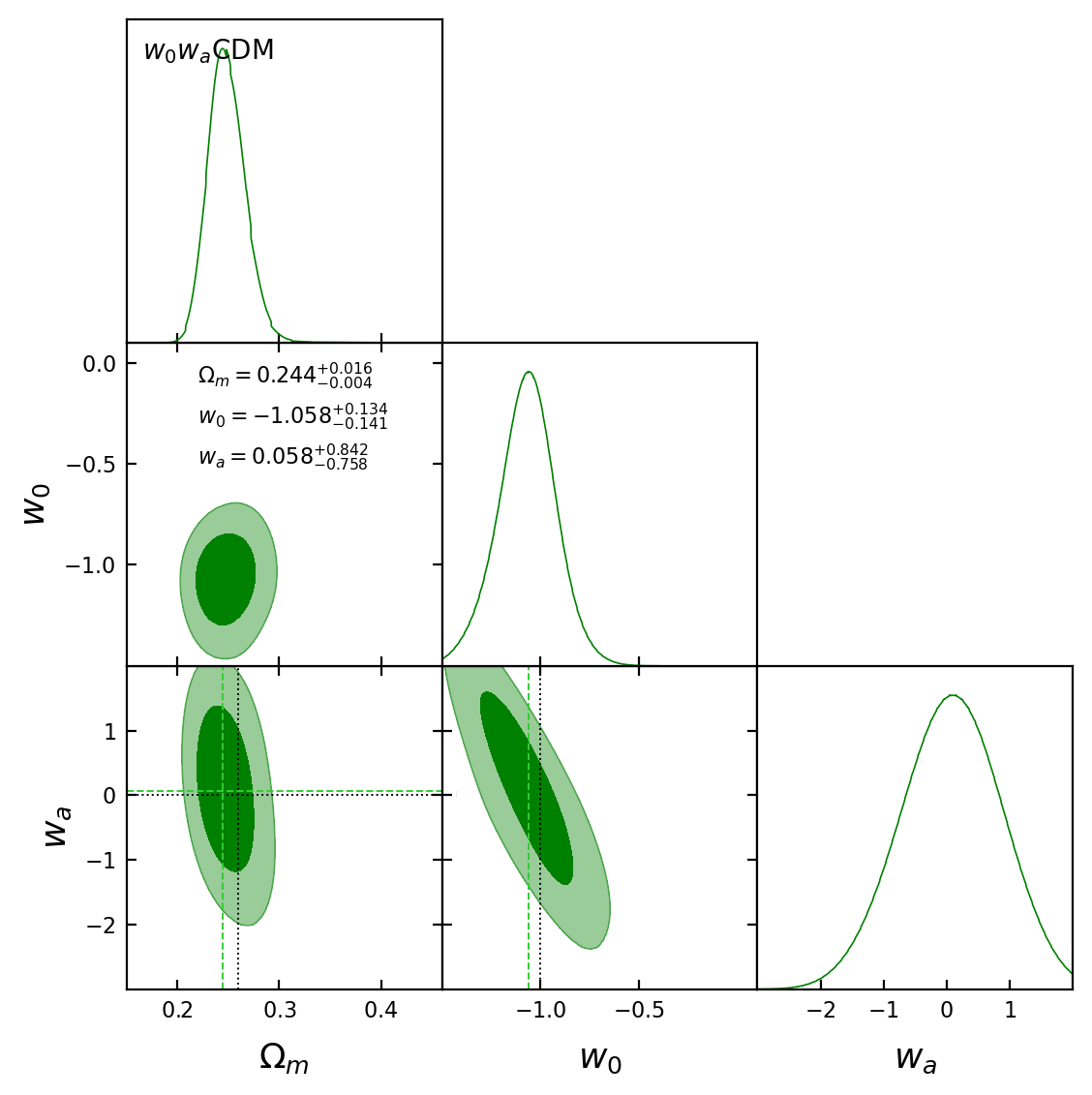}}
   \caption{Same as Figure \ref{fig:result_APBAO}, but from mocks as the input–output test. Dotted lines locate the input cosmology and dashed lines correspond to the output cosmology.}
    \label{fig:apmock}
\end{figure*}

We begin by deriving the galaxy redshifts under the true cosmology. For each cosmological model adopted, we then convert the redshifts of galaxies in the mock samples into comoving distances and measure the normalized correlation function, $\xi(\Omega_m, w, z)$, in six redshift bins. Following the likelihood analysis outlined in Section \ref{sec:likelihood}, we derive constraints on the cosmological parameters. This procedure is repeated across different mock realizations, and the resulting probability distribution functions (PDFs) are averaged, as shown in Figure \ref{fig:apmock}. We employ this approach to quantify potential systematic biases within our analysis pipeline. 

Figure \ref{fig:apmock} reveals a parameter degeneracy direction consistent with the AP test contours in Figure \ref{fig:result_APBAO}. Furthermore, the input cosmology is successfully recovered, albeit with minor shifts in the parameters: $\delta\Omega_m = -0.016$, $\delta w_0 = -0.058$, and $\delta w_a = 0.058$. In \S\ref{sec:result}, we apply these systematic parameter offsets as corrections to shift the observed PDF of ($\Omega_m$, $w_0$, $w_a$) within the parameter space.

\end{document}